\documentclass[]{spie}  

\usepackage{amsmath, amsfonts, amssymb, braket, dsfont}
\usepackage{graphicx}
\usepackage[colorlinks=true, allcolors=blue]{hyperref}
\usepackage{tikz}
\usetikzlibrary{quantikz}

\graphicspath{{figures/}}

\newcommand{\washington}{ibmq\_washington}
\newcommand{\mumbai}{ibmq\_mumbai}

\title{Assessing the Stability of Noisy Quantum Computation}

\author[a]{Samudra Dasgupta}
\author[b]{Travis S. Humble}
\affil[a]{University of Tennessee, Knoxville, Tennessee USA}
\affil[b]{Oak Ridge National Laboratory, Oak Ridge, Tennessee USA}

\authorinfo{Further author information: (Send correspondence to T.S.H.)\\T.S.H.: E-mail: humblets@ornl.gov\\  S.D.: E-mail: sdasgup3@vols.utk.edu}

\begin{document} 
\maketitle
\begin{abstract}
Quantum computation has made considerable progress in the last decade with multiple emerging technologies providing proof-of-principle experimental demonstrations of such calculations. However, these experimental demonstrations of quantum computation face technical challenges due to the noise and errors that arise from imperfect implementation of the technology. Here, we frame the concepts of computational accuracy, result reproducibility, device reliability and program stability in the context of quantum computation. In particular, we provide intuitive definitions for these concepts that lead to operationally meaningful bounds on program output. Our assessment highlights the continuing need for statistical analyses of quantum computing program to increase our confidence in the burgeoning field of quantum information science.
\end{abstract}


\section{Introduction}\label{sec:intro}
The stability of a computing system is an important concern for assessing computational performance. 
System stability plays an essential role in qualifying the reproducibility of scientific results as well the validation, verification, and quantification of uncertainty. For quantum computing, the underlying quantum device stability  also represents an essential concern for both reliable and reproducible results. Unique sources of quantum device instability include the varying quality of the quantum register implementation, which suffers from non-uniform spontaneous decay, energy loss to the environment, cross-talk, the sensitivity of gate operations for initialization, measurement, and unitary operations to imprecise control pulses as well as fluctuations in the thermodynamic controls. In the presence of inhomogeneous and non-stationary noise processes, an unstable quantum device presents a unique challenge for useful computation as it becomes difficult to attribute errors and quantify uncertainty. While there have been many successful demonstrations of accurate quantum computing, these are generally point solutions using highly calibrated and tuned devices. Thus, little attention has been paid to whether such results are reproducible across varying technologies or instances of a given technology. The resulting notion of stability tests the preciseness of repeating a quantum computation, for example, within a defined tolerance, as well as the reproducibility of the results as measured by the similarity between independent outcome distributions. 
\par 
In this study, we investigate how to assess the stability of a quantum device and link this metric to result reproducibility and device reliability. The notions of accuracy, reproducibility, reliability and stability are related yet distinct. We use the term accuracy to quantify how close the experimentally realized output of a quantum program matches the known, correct output \cite{lilly2020modeling}. Reproducibility is used to quantify how close an instance of a program is to a previously executed instance on the same or potentially different device \cite{dasgupta2021reproducibility,dasgupta2022characterizing}. Reliability studies how select parameters characterizing the device components, such as gate fidelities, deviate from expectation, while stability quantifies whether the program behavior stays bounded in the presence of fluctuations due to noise \cite{dasgupta2021stability}. In the following sections, we define and quantify these notions and, with the help of depolarizing channel as a running example, explore their behaviors for noisy quantum circuits.
\section{Theory}\label{sec:theory}
Consider an idealized quantum computer 
composed of an $n$-qubit register that encodes a $2^{n}$-dimensional Hilbert space $\mathbb{C}^{2\otimes n}$ and a set of operations that transform the register. Projective operations initialize the state of the register, e.g., in the computational basis state $\ket{\psi} = \ket{0}^{\otimes n}$, while a unitary operation $\mathcal{U}$, aka a gate, transforms the state as $\ket{\psi} \rightarrow \mathcal{U}\ket{\psi}$. Additional projective operations implement measurement of the register labeled by the $n$-bit string \textit{s}, where $s \in \{0, 1\}^{\otimes n}$, and the probability to observe this outcome is $\textrm{Pr}(s) = |\braket{s|\psi}|^2$. 
Practical efforts to realize a quantum computer introduce many additional physical processes, identified here as noise, that complicate the operational description above \cite{kliesch2021theory, ferracin2021experimental, coveney2021reliability, blume2010optimal}. 
In all cases, an important question is to understand whether the resulting computational output is accurate, reproducible and stable. We next consider these notions in the presence of quantum channels that model noisy operations. 
\subsection{Noisy circuit model}\label{sec:model}
Consider the initial density matrix representing the state of a quantum register as $\rho_{0}$ and let the noiseless application of a circuit unitary $U_C$ prepare the state $\rho = U_{C} \rho_{0} {U_{C}}^\dagger$. 
By comparison, let the action of a noisy circuit on this state be modeled as $\rho' = \mathcal{E}(\rho)= {\textstyle\sum}_{k} M_k \rho M_k^\dagger$, where $\mathcal{E}$ is a channel operator expressed in terms of Kraus operators ($M_k$). In practice, such a simple model may be replaced by more complex, constructive models for the individual gate operations \cite{lilly2020modeling}.
In the orthonormal computational basis $\{\ket{i}\}_{i=0}^{N-1}$ with $N=2^n$, a corresponding projective measurement is modeled as $\Pi_i = \ket{i}\bra{i}\textrm{ for }{i=0\textrm{ to }N-1}$. The noiseless circuits yield the probability for the $i$-th outcome as $q_i = \textrm{Tr}[\Pi_i \rho]$, while the noisy circuit model yields the probability
\begin{equation}
   p_i = \textrm{Tr}[\Pi_i^\dagger \Pi_i \rho'] = \textrm{Tr}[\Pi_{i}^{\dagger}\sum_{k}{M_{k}\rho M_{k}^{\dagger}}\Pi_{i}]
\end{equation}
The latter models may also be extended to consider noisy readout channels \cite{smith2021qubit}.
\par 
Differences in the probabilities influence the observed measurement results as well as the derived expectation values. For example, consider the case of a register with $n=1$ qubits in the presence of depolarizing noise. The latter channel operator is characterized by a noise parameter $e$ for which the Kraus operators $M_{k} \in \{\sqrt{1-e}I, \sqrt{e}X, \sqrt{e}Y, \sqrt{e}Z \}$ yield  
\begin{align}
\rho' = \mathcal{E}_e(\rho) =& (1-e) \rho + \frac{e}{3}X \rho X+ \frac{e}{3}Y \rho Y+ \frac{e}{3} Z \rho Z
\end{align}
Assuming $\rho = \ket{\psi}\bra{\psi}$ with $\ket{\psi} = \alpha \ket{0} + \beta \ket{1}$, then $q_0 = |\alpha|^2$ and $q_{1} = |\beta|^2$, while $p_{0} = q_{0} - 2e(1-2|\beta|^2)/3$ and $p_1 = q_1 + 2e(1-2|\beta|^2)/3$ and the noisy expectation value of the diagonal observable $Z$ is $\braket{Z_{noisy}} = (-1) p_{0} + (+1) p_1 = (2|\beta|^2-1)\left(1-\frac{4}{3}e\right)$. 
\subsection{Computational accuracy}\label{sec:accuracy}
It is apparent above that noise induces discrepancies in the probabilities for measurement outcomes and that these errors will make the resulting observables deviate from their expected behavior. We quantify the difference in the noisy and noiseless probability distributions, $p$ and $q$, in terms of the well-known Hellinger distance \cite{}. The Hellinger distance between $p$ and the reference distribution $q$ is defined as $d_{\mathcal{H}}(p, q) = \sqrt{1-\textrm{BC}(p, q)}$ where the Bhattacharyya coefficient is given by  $\textrm{BC}(p, q) = \sum_{i}{\sqrt{p_i q_i}}$ with $p_{i}, q_{i}$ the $i$-th discrete output of these distributions. It is notable that the Hellinger distance vanishes when the distributions are identical and grows to unity as the distributions become completely disjoint. Using our expression for noisy circuit execution per the model discussed in the previous section, the Hellinger distance may be expressed as
\begin{equation}
    d_{\mathcal{H}} = \sqrt{1-\sum\limits_i\sqrt{\sum_{k}{ \textrm{Tr}[\Pi_{i}M_{k} \rho M_{k}^{\dagger}] \textrm{Tr}[\Pi_{i}\rho]}}}
\end{equation}
Given this measure of similarity, we now say the distribution $p$ is $\epsilon-$accurate if $d_{\mathcal{H}} \leq \epsilon$. Checking this condition for a given error tolerance $\epsilon$ requires a simple testing after computing the Hellinger distance.
\par 
However, the above definition of accuracy requires a priori knowledge of the noiseless reference distribution $q$, and this may be an impractical requirement when testing the accuracy of noisy quantum circuits whose noiseless distribution of outcomes is unknown, e.g., for large problem sizes in the realm of quantum advantage. In such cases, accuracy of the measurement probabilities may be verified indirectly. In particular, we can define the error in a computed observable as
\begin{equation}
d_{O} = |\braket{O_{noisy}} - \braket{O_{noiseless}}|
\end{equation}
and the $\epsilon-$accuracy condition can be stated as:
\begin{align}
&d_{O} \leq \epsilon\\
\Rightarrow &|\sum\limits_m m\left( p(m) - q(m)\right)| \leq \epsilon\\
\Rightarrow &| \sum\limits_m m \textrm{Tr}\left[\Pi_m \mathcal{E}_e(\rho) -\Pi_m \rho\right]| \leq \epsilon 
\end{align}
where $\mathcal{E}_e(\cdot)$ denotes the noise channel for the parameter instance $e$. Here, we  define $q(m) = \textrm{Tr}[\Pi_m \rho]$ as the probability of observing eigenvalue $m$ of the observable $O$ when projecting onto the corresponding $m$-th eigenstate by the measurement $\Pi_m$, and $p(m) = \textrm{Tr}[\Pi_m \mathcal{E}_e(\rho) ]$ as the same for the noisy channel $\mathcal{E}_e(\cdot)$. 
\par 
Consider our single-qubit working example in the presence of depolarizing noise discussed in the previous section. Then, the Hellinger distance is given as
\begin{align}
    d_{\mathcal{H}} = \left(1-|\alpha|^2\sqrt{1-\frac{2e}{3}\left(1-\left|\frac{\beta}{\alpha}\right|^2\right)}-|\beta|^2\sqrt{1-\frac{2e}{3}\left(1-\left|\frac{\alpha}{\beta}\right|^2\right)}\right)^{1/2}
\end{align}
while the error in our diagonal observable becomes 
\begin{equation}
d_{O} = |\braket{Z_{noisy}}-\braket{Z_{noiseless}}| = \left|\frac{4e ( 1-2 |\beta|^{2} ) }{3}\right|
\end{equation}
Assuming a desired computational accuracy that is bounded by error $\epsilon$, this places an upper bound on the depolarizing channel parameter as $e \leq \frac{3\epsilon}{4}$
%
\begin{figure*}
  \centering
  \begin{tabular}{ c @{\hspace{40pt}} c }
      \includegraphics[width=.4\columnwidth]{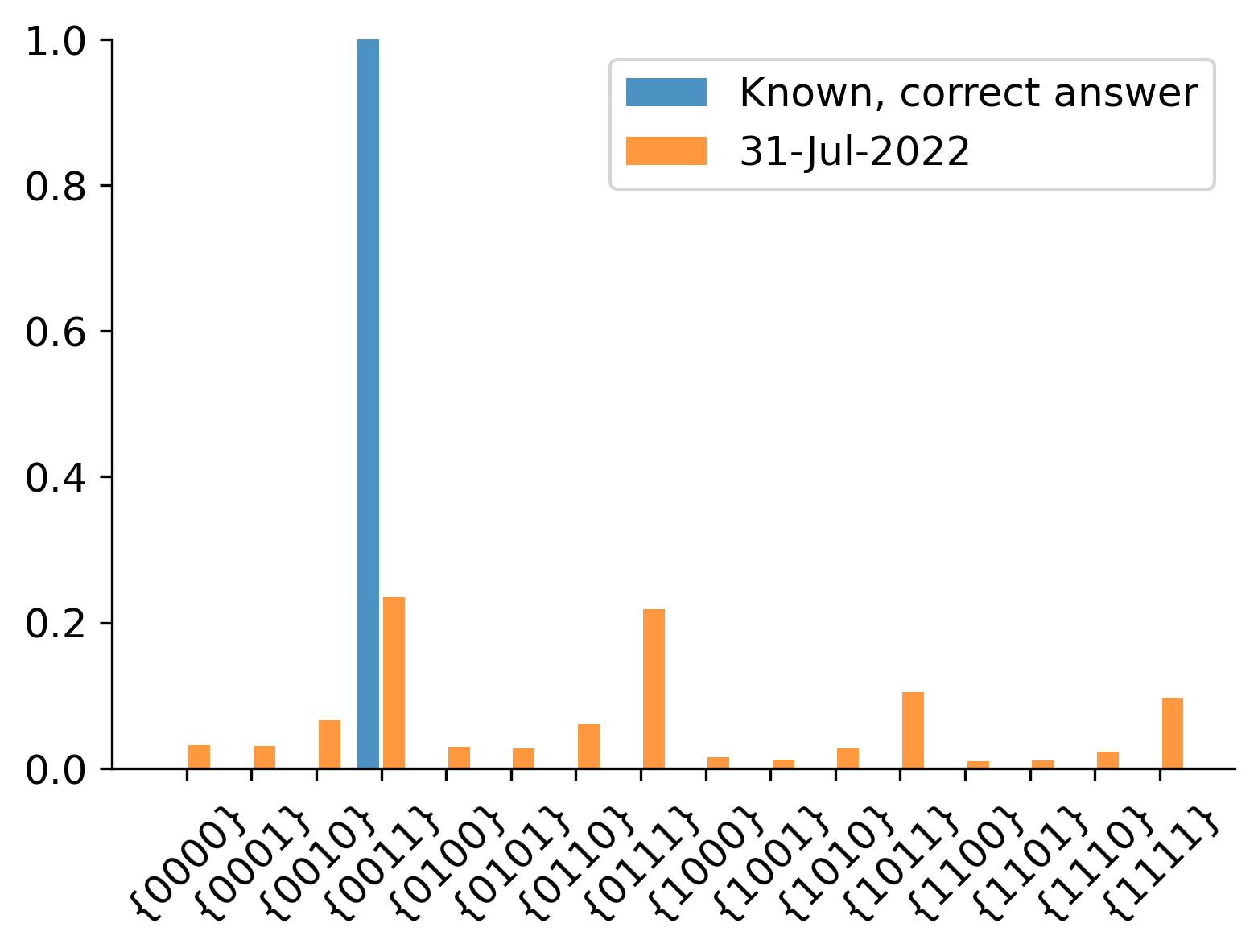} &
    \includegraphics[width=.4\columnwidth]{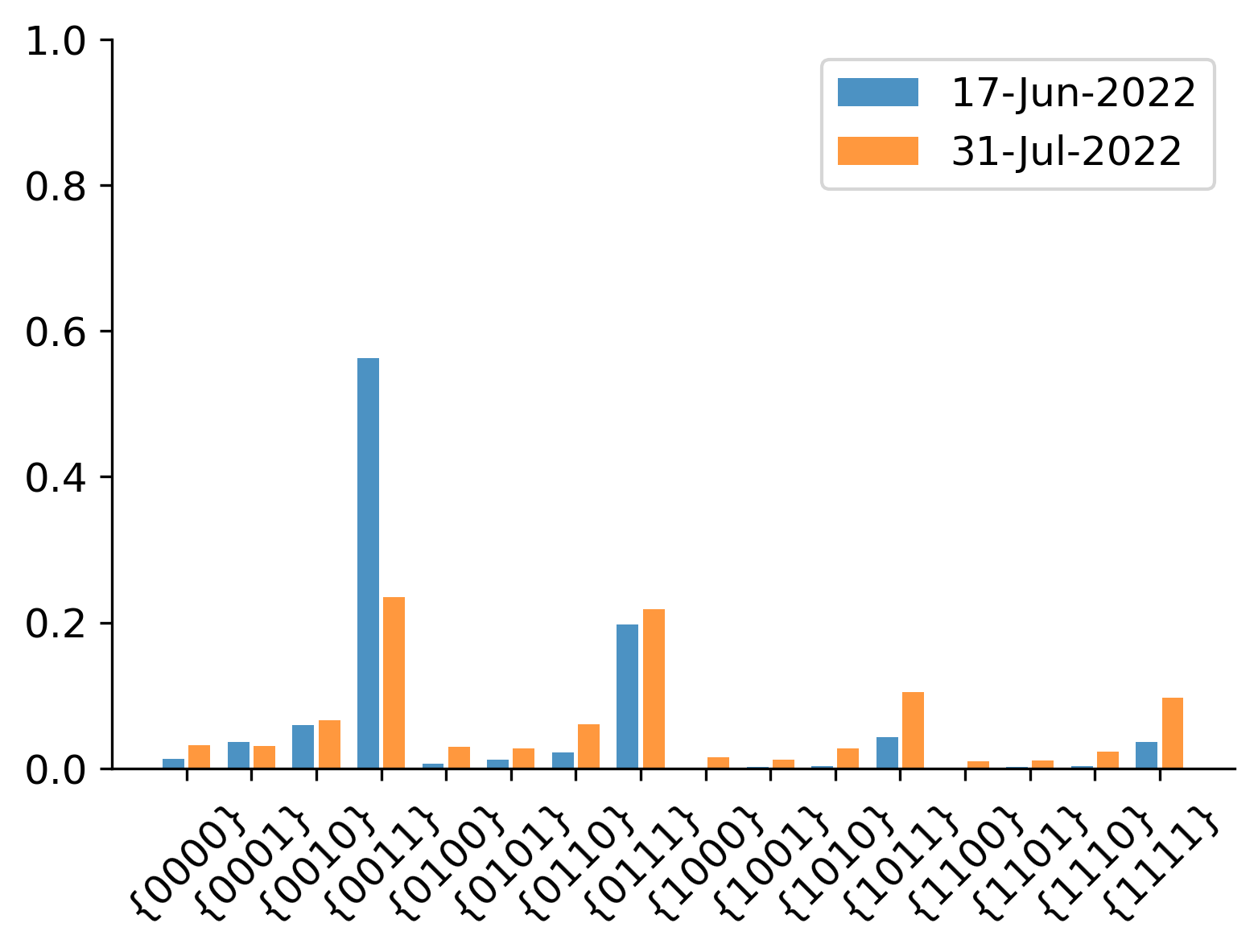} \\
    \small (a) &
      \small (b)
  \end{tabular}

  \medskip

  \caption{Computational accuracy and result reproducibility: (a) Expected versus experimentally observed measurement distributions for a 4-qubit Bernstein-Vazirani program executed on the 27-qubit transmon processor \mumbai. (b) Experimentally observed distributions for the same problem for 31-July-2022 vs 17-June-2022.
  }
\label{fig:accuracy_repr}
\end{figure*}
%
\subsection{Result reproducibility}\label{sec:repr}
In the previous section, we elaborated on the meaning of accuracy in the context of quantum computing, for which the output from a program instance can be tested to satisfy as desired accuracy conditions. However, such tests are instance specific and do not address the question of whether the program is reproducible, in the sense that repeated executions will generate similar results. The latter statistical concept requires consideration of program output across multiple instances, which may include execution at different times or on different devices [e.g. see Fig.~\ref{fig:accuracy_repr}]. In particular, the channel parameters, denoted by $e$, may be different across such instances. We therefore consider the Hellinger distance of the program output with respect to multiple execution instances and pose the confidence internal denoted by 
$\textrm{Pr}(d_{\mathcal{H}}^{e} \leq \epsilon) \geq 1 - \delta$, where $\delta$ denotes the desired confidence level (e.g., 0.05) and $e$ signifies the instance of channel noise. With respect to error in an observable, we may similarly pose the reproducibility condition as 
\begin{align}
\textrm{Pr}(d_{O}^{e} \leq \epsilon) &\geq 1 - \delta
\end{align}
where $d_{O}^{e} =\left|\braket{O^{ideal}}-\braket{O^{noisy}_{e}}\right|$
and $\braket{O^{noisy}_e}$ is a random variable due to the presence of both shot noise as well as the randomness of $e$.
This reproducibility condition may be used to derive a bound on the error model. For example, in our working demonstration, we find that
\begin{align}
\textrm{Pr}\left(  e \leq \left|\frac{3\epsilon}{4(1-2|\beta|^2)} \right|\right) >& 1 - \delta
\end{align}
If the parameter $e$ is drawn from a probability distribution $p(e)$ that follows an exponential distribution with mean $1/\lambda$, i.e.,
\begin{align}
p_{\lambda}(e) = \lambda \exp^{-\lambda e},
\end{align}
then the reproducibility condition requires $\lambda
\geq {4|\log \delta |} / {3\epsilon}$. 
\subsection{Device reliability}\label{sec:rel}
We call a device reliable if the characteristic parameters describing its quantum noise channels (such as $e$ and $\lambda$ in our working example) remain nearly stationary across multiple instances of program execution. These parameters are dependent upon the model selected to characterize the device and we have previously proposed DiVincenzo's criteria to generate a minimal set \cite{divincenzo2000physical}. Those characteristic metrics include the register capacity, initialization fidelity, gate fidelity, 
duty cycle, which we define as the ratio of register coherence time to gate duration, and register  addressability \cite{dasgupta2021stability}. The latter is quantified in terms of the mutual information between register elements. These metrics are often used as characteristic parameters that inform more explicit noise models for quantum devices \cite{lilly2020modeling}. 
\par
\begin{figure*}
  \centering
  \begin{tabular}{ c @{\hspace{40pt}} c }
      \includegraphics[width=.4\columnwidth]{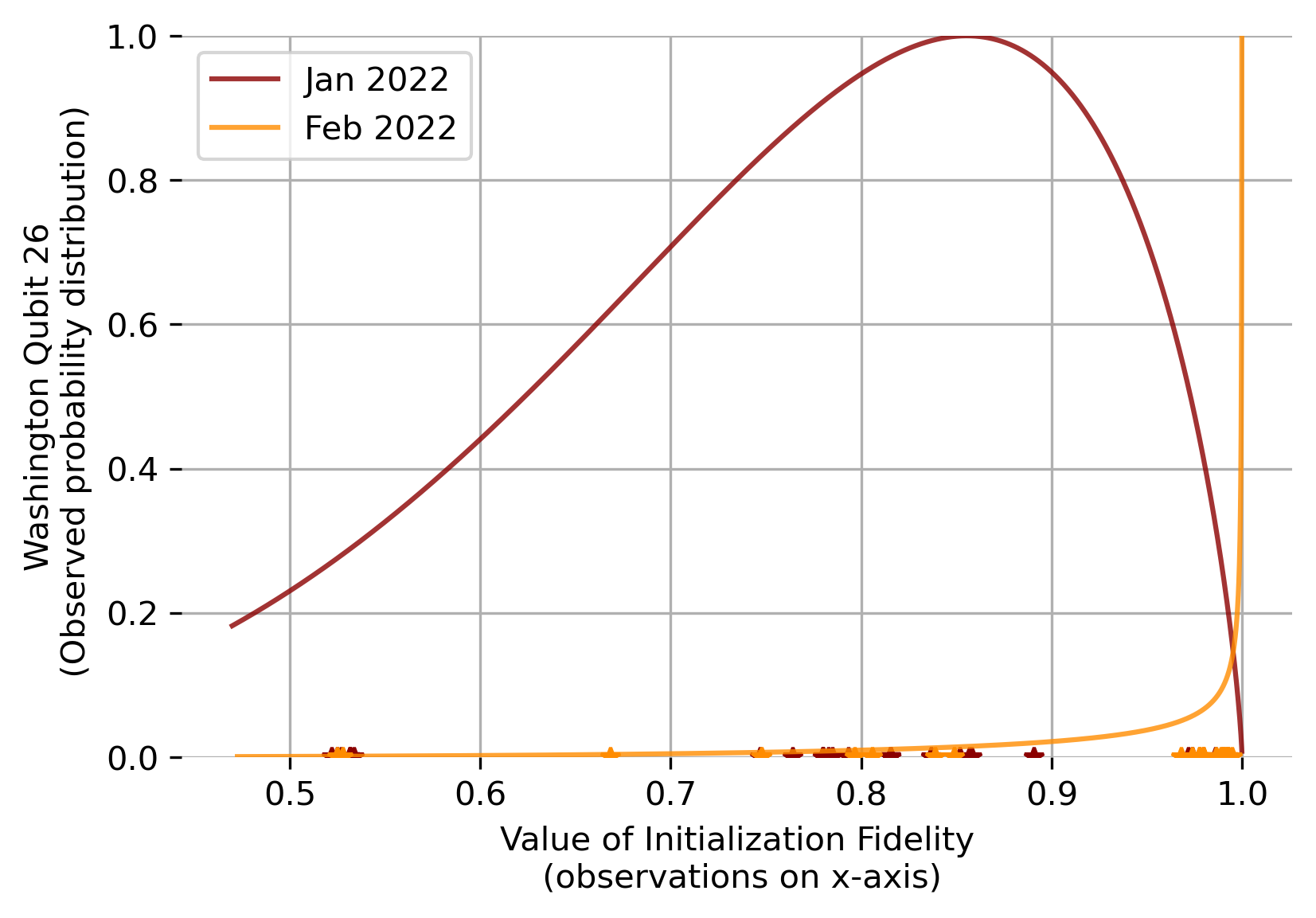} &
    \includegraphics[width=.4\columnwidth]{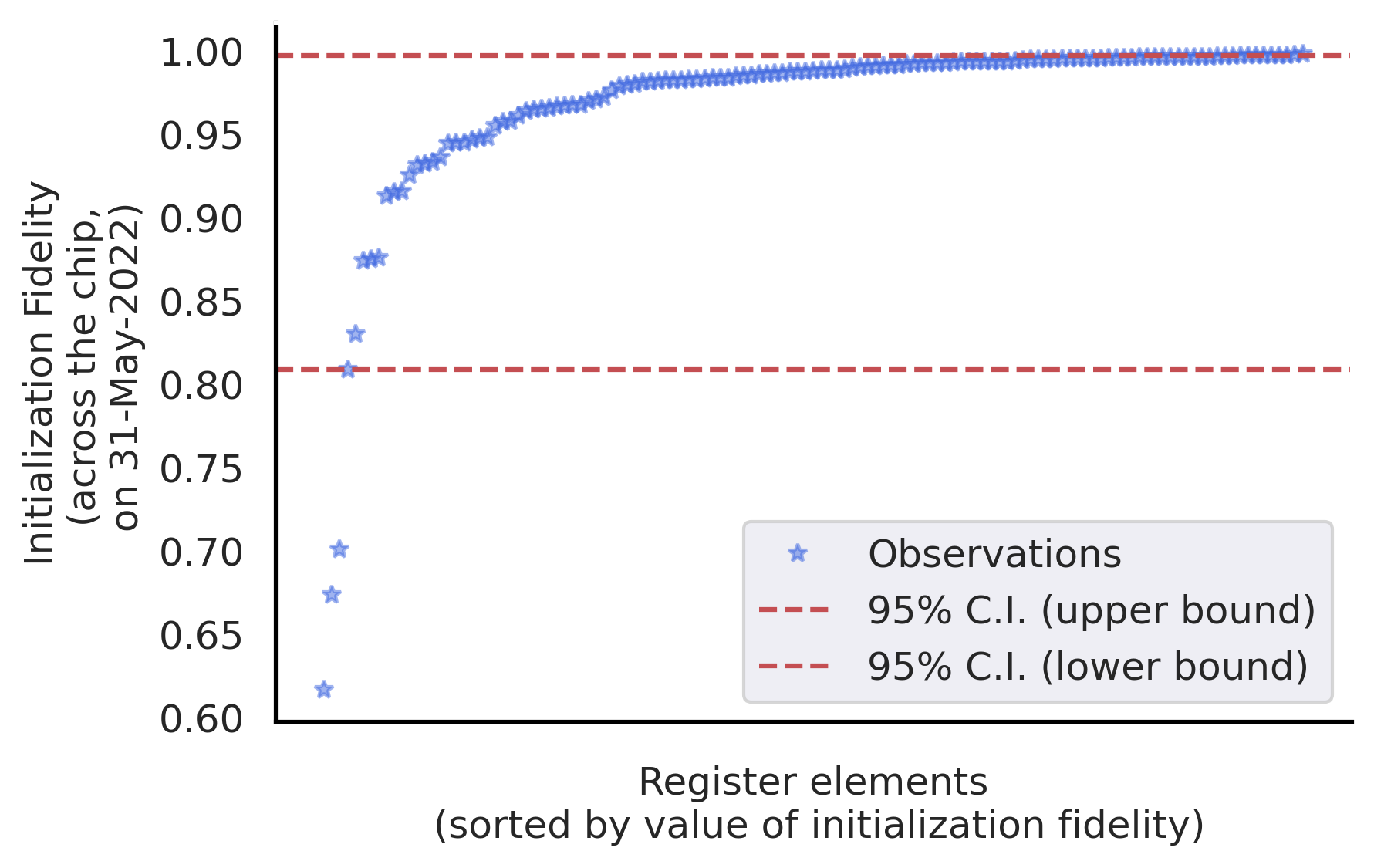} \\
    \small (a) &
      \small (b)
  \end{tabular}
  \medskip
  \caption{Device reliability measures a device's consistency with respect to characteristic parameters. (a) Wide variance seen in the distribution of Initialization Fidelity, indicating an unstable readout channel. Data shown for qubit \#26. (b) Wide spatial variation seen across the 127 qubits of \washington. Data shown for 31-May-2022.
}

\label{fig:q26_washington_dec_may2022}
\end{figure*}

 
For example, variations in the distribution of the initialization fidelity for the \washington ~device are shown in Fig.~\ref{fig:q26_washington_dec_may2022}. Here, the fidelity metric observed during two different periods of time has been fit to the beta distribution, indicating stark differences of the behavior during these two execution windows.
The noise parameter characterizing the readout error on May 31, 2022 fluctuates significantly across the device as well as time. Thus, the assumption that a NISQ noise channel is stationary does not hold up to experimental scrutiny. However, noise characterizations are assumed to be constant in current research when applying error mitigation. Thus, a natural question arises as to how sensitive is the output to non-stationary noise processes.
\subsection{Output stability}\label{sec:stability}
In the last section, we discussed how today's noisy devices used for quantum computing can be non-stationarity which makes the analysis using quantum error channel formalism more challenging. In particular, the distribution of error parameters shows a time-dependence:
\begin{equation}
p(e) = p(e;t)
\end{equation}
For example, in Fig.~\ref{fig:q26_washington_dec_may2022}, the readout error channels exhibit non-stationarity.

The presence of such time-varying parameter distribution makes the quantum error channel $\mathcal{E}_e(\cdot)$ itself non-stationary.

This in turn can make the result obtained from executing a quantum program a non-stationary stochastic process e.g. the result may exhibit a drifting mean or time-varying error bars. We call this the problem of output stability (or program stability). 

Thus, stability investigates the program execution sensitivity to the degree of non-stationarity of the error channel. Note that stationary error channels will produce stable output but it may still be irreproducible and inaccurate (recall that accuracy concerns itself with correctness while reproducibility concerns itself with ability to replicate a previously obtained result).


Let p(e;t) denote the distribution of the error parameter (possibly non-stationary) characterizing the quantum error channel $\mathcal{E}_e(\cdot)$.
Then the expected output of a noisy quantum computation at time t:
\begin{equation}
\braket{O_e(t)} = \int\limits_e p(e;t) O_e de
\end{equation}
where $O_e$ is the output when the error channel $\mathcal{E}_e(\cdot)$ is characterized by the fixed error parameter $e$. Note that the output $O_e$ can be either a real number such as the mean of an observable or a distribution across discrete states.

The sensitivity of the expected output of the execution of a quantum program at two different times $t_1$ and $t_2$ is then: 
\begin{equation}
\delta_t \braket{O_e} = \int\limits_e \braket{O_e} \left\{ p(e; t_1) - p(e; t_2)\right\} de
\end{equation}
Note that this captures the variability attributable to the non-stationarity of the quantum error channel only and not due to the Born rule or the randomness of the error parameter, which is what we desire.

The $\epsilon-$stability condition can then be stated as:
\begin{equation}
\left| 
\int\limits_e \braket{O_e} \left\{ p(e; t_1) - p(e; t_2)\right\} de
\right| \leq \epsilon
\end{equation}
It is important to note that the stability condition is defined with respect to a time-interval $\delta t = (t_1, t_2)$.
The analysis of the $\epsilon-$stability condition can help us bound the degree of channel non-stationarity permissible within this time-interval.

In fact, a general bound on the degree of non-stationarity for $e$ (in terms of the Hellinger distance) can be given by:
\begin{align}
d_{\mathcal{H}}^e \leq  \sqrt{1-\sqrt{2\phi^2-1}} \textrm{ where, } \phi = \frac{\epsilon}{2\sqrt{2}}\left[\int \left|\frac{\braket{O_e}}{\braket{O}}\right|^2 de\right]^{-1/2}
\end{align}

\begin{proof}
We want to bound the degree of input non-stationarity of $e$ in terms of the stability tolerance $\epsilon$.
To measure the degree of input non-stationarity, we use the Hellinger distance $d_{\mathcal{H}}^e$ between $p(e; t_1)$ and $p(e; t_2)$
For convenience, define:
\begin{align*}
p =& p(e;t_1)\\
q =& p(e;t_2)\\
f =& \braket{O_e}\\
M =& \braket{O^{\textrm{ideal}}}
\end{align*}
Then,
\begin{align}
\delta_t \braket{O_e} &= \left| \int f(p-q)de \right|\\
&= \left| \int f(\sqrt{p}-\sqrt{q})(\sqrt{p}+\sqrt{q})de \right|\\
&\leq \sqrt{\int f^2 de}\sqrt{\int (\sqrt{p}-\sqrt{q})^2 de}\sqrt{\int (\sqrt{p}+\sqrt{q})^2 de} \textrm{ ( from Cauchy-Schwarz inequality) }\\
&= 2\sqrt{2}cd \sqrt{1-\frac{d^2}{2}}
\end{align}

where $c = \sqrt{\int f^2 de}$ and $d = d_{\mathcal{H}}^e(p,q) =$ Hellinger distance between $p$ and $q$.

Using the stability condition, this yields:
\begin{align}
\frac{2\sqrt{2}cd \sqrt{1-\frac{d^2}{2}}}{M} &\leq \epsilon\\
\Rightarrow d \sqrt{1-\frac{d^2}{2}} &\leq \phi^2 \textrm{ where } \phi = \frac{\epsilon M}{2\sqrt{2}c} = \frac{\epsilon}{2\sqrt{2}}\left[\int \left|\frac{\braket{O_e}}{\braket{O}}\right|^2 de\right]^{-1/2}\\
\Rightarrow d &\leq \sqrt{1-\sqrt{2\phi^2-1}}
\end{align}\end{proof}

For our running example of a depolarizing channel, we get:
\begin{equation}
d_{\mathcal{H}}^{e}(p(e;t_1), p(e;t_2)) \leq \sqrt{1-\sqrt{2\phi^2-1}}
\end{equation}
where $\phi = \frac{3\sqrt{3}\epsilon}{14\sqrt{2}}$.

Our framework enables the question of time-scale for which calibrations are necessary to be addressed. Suppose the rate of non-stationary dynamics increases with time. This process can then be modeled by assuming that the spread of the error parameter $e$ characterizing the channel also increases with time. Since the spread (or variance) of an exponential distribution $\propto \lambda^{-2}$, we can then model this situation assuming channel-drift $\lambda(t) = \lambda_0 -\eta \delta t$ i.e. $\lambda$ decreases at a rate $\eta$. The time-scale at which the device transitions from stable to unstable (in the absence of re-calibration) can be estimated as:
\begin{align}
d_{\mathcal{H}}^{e}\left[p(e;t), p(e;t+\delta t)\right] &\leq \sqrt{1-\sqrt{2\phi^2-1}} \\
\Rightarrow \sqrt{1-\frac{\lambda_0(\lambda_0-\eta\delta t)}{2\lambda_0-\eta\delta t}} &\leq \sqrt{1-\sqrt{2\phi^2-1}}\\
\Rightarrow \delta t_{\textrm{stable}} &\leq \frac{\lambda_0 2\sqrt{2}\sqrt{1-\sqrt{2\phi^2-1}}}{\eta}
\end{align}
where $\phi = \frac{3\sqrt{3}\epsilon}{14\sqrt{2}}$. Thus, in presence of this channel-drift model, we expect a stable device to remain stable until $t=\delta t_{stable}$. This gives an estimate of the frequency of device calibration required to meet the stability condition.

In this section, we discussed the problem of program instability which has severe implications for the interpretation of the results obtained from execution of a quantum program on a noisy device. For example, the results may need to be corrected for bias (such as time-drift) and error mitigation schemes may require adaptive (Bayesian) channel estimation (which maybe expensive due to the natural tension between the number of statistical samples required for channel estimation vs the time-scales at which the non-stationarity manifests).
\section{Conclusions}\label{sec:conclusions}
Unstable fluctuations in device parameters represent a significant concern for the reproducibility of NISQ computing demonstrations. 
Many current experimental demonstrations rely on quantum circuits calibrated immediately prior to program execution and tuned during run-time. 
While this approach is successful for singular demonstrations, the resulting circuits and calibrations are implicitly dependent on the device parameters, which fluctuate significantly over time, across the chip and, technology. Unstable devices are not suitable for error attribution or uncertainty quantification (let alone producing reliable results). Most of the research till date has focused on accuracy but little has been done on the reproducibility and stability of quantum computers.
In this study, we contribute a first step addressing this issue and define computational accuracy, result reproducibility and device reliability and discuss how they are linked using a framework for stability of quantum information. Without additional efforts to make current experimental results reproducible, the knowledge and insights gained from today's burgeoning field of quantum computer research may be undercut by low confidence in the reported results.
%
\section*{ACKNOWLEDGMENTS}
This work is supported by the U.~S.~Department of Energy (DOE), Office of Science, National Quantum Information Science Research Centers, Quantum Science Center and the Advanced Scientific Computing Research, Advanced Research for Quantum Computing programs. This research used computing resources of the Oak Ridge Leadership Computing Facility, which is a DOE Office of Science User Facility supported under Contract DE-AC05-00OR22725. The manuscript is authored by UT-Battelle, LLC under Contract No.~DE-AC05-00OR22725 with the U.S. Department of Energy. The U.S.~Government retains for itself, and others acting on its behalf, a paid-up nonexclusive, irrevocable worldwide license in said article to 
reproduce, prepare derivative works, distribute copies to the public, and perform publicly and display publicly, by or on behalf of the Government.  
The Department of Energy will provide public access to these results of federally sponsored research in accordance with the DOE Public Access Plan. 
http://energy.gov/downloads/doe-public-access-plan.

\bibliographystyle{spiebib} 
\bibliography{references.bib}

\begin{thebibliography}{10}

\bibitem{lilly2020modeling}
Lilly, M.~N. and Humble, T.~S., ``Modeling noisy quantum circuits using
  experimental characterization,'' {\em arXiv preprint arXiv:2001.08653}
  (2020).

\bibitem{dasgupta2021reproducibility}
Dasgupta, S. and Humble, T.~S., ``Reproducibility in quantum computing,'' in
  [{\em 2021 IEEE Computer Society Annual Symposium on VLSI
  (ISVLSI)}{\nolinebreak\hspace{0.1em}]},   458--461, IEEE (2021).

\bibitem{dasgupta2022characterizing}
Dasgupta, S. and Humble, T.~S., ``Characterizing the reproducibility of noisy
  quantum circuits,'' {\em Entropy}~{\bf 24}(2),  244 (2022).

\bibitem{dasgupta2021stability}
Dasgupta, S. and Humble, T.~S., ``Stability of noisy quantum computing
  devices,'' {\em arXiv preprint arXiv:2105.09472}  (2021).

\bibitem{kliesch2021theory}
Kliesch, M. and Roth, I., ``Theory of quantum system certification,'' {\em PRX
  Quantum}~{\bf 2}(1),  010201 (2021).

\bibitem{ferracin2021experimental}
Ferracin, S., Merkel, S.~T., McKay, D., and Datta, A., ``Experimental
  accreditation of outputs of noisy quantum computers,'' {\em arXiv preprint
  arXiv:2103.06603}  (2021).

\bibitem{coveney2021reliability}
Coveney, P.~V., Groen, D., and Hoekstra, A.~G., ``Reliability and
  reproducibility in computational science: implementing validation,
  verification and uncertainty quantification in silico,'' {\em Philosophical
  Transactions of the Royal Society A}~{\bf 379}(2197),  20200409 (2021).

\bibitem{blume2010optimal}
Blume-Kohout, R., ``Optimal, reliable estimation of quantum states,'' {\em New
  Journal of Physics}~{\bf 12}(4),  043034 (2010).

\bibitem{smith2021qubit}
Smith, A.~W., Khosla, K.~E., Self, C.~N., and Kim, M., ``Qubit readout error
  mitigation with bit-flip averaging,'' {\em Science Advances}~{\bf 7}(47),
  eabi8009 (2021).

\bibitem{divincenzo2000physical}
DiVincenzo, D.~P., ``The physical implementation of quantum computation,'' {\em
  Fortschritte der Physik: Progress of Physics}~{\bf 48}(9-11),  771--783
  (2000).

\end{thebibliography}
%
\end{document}